\documentclass[10pt,english,journal]{IEEEtran}
\usepackage{hyperref}
\pdfoutput=1
\usepackage[T1]{fontenc}
\usepackage[latin9]{inputenc}
\usepackage{geometry}
\usepackage{amsmath}
\usepackage{xpatch}
\usepackage{amssymb}
\usepackage{esint}
\usepackage{mathtools}
\usepackage{amsthm}
\usepackage{nicefrac}
\usepackage{gensymb}
\usepackage{bigints}
\usepackage{mathtools}
\usepackage{blkarray, bigstrut}
\usepackage{physics}
\usepackage{calligra}
\usepackage{graphicx}
\usepackage{epstopdf}
\usepackage{dsfont}
\usepackage{array,ragged2e}
\usepackage{enumitem}
\usepackage{etoolbox}
\usepackage{babel}
\usepackage[nopar]{lipsum}
\usepackage{psfrag}
\usepackage[ruled,norelsize]{algorithm2e}
\usepackage{algorithmic}
\usepackage{algorithm2e}
\usepackage{balance}
\usepackage{float}
\usepackage[caption = false]{subfig}
\SetKw{KwBy}{by}

\makeatletter
\newcommand{\removelatexerror}{\let\@latex@error\@gobble}
\makeatother

\newtheoremstyle{plain}
  {\topsep}   
  {\topsep}   
  {\itshape}  
  {0pt}       
  {\bfseries} 
  {.}         
  {5pt plus 1pt minus 1pt} 
  {\thmname{#1}\thmnumber{ #2} \textnormal{(\thmnote{#3})}} 

\xpatchcmd{\proof}{\hskip\labelsep}{\hskip5\labelsep}{}{}  
\makeatletter
\xpatchcmd{\proof}{\@addpunct{.}}{\@addpunct{:}}{}{}
\makeatother

\renewcommand\[{\begin{equation}}
\renewcommand\]{\end{equation}} 
\newcounter{MYtempeqncnt}
\usepackage[usenames,dvipsnames,table]{xcolor}
\usepackage{listings}
\usepackage{fancyvrb}
\usepackage{framed}

\usepackage{courier}

\definecolor{dkgreen}{rgb}{0,0.3,0}
\definecolor{gray}{rgb}{0.5,0.5,0.5}

\usepackage{tcolorbox}
\usepackage{tabularx}
\usepackage{array}
\usepackage{colortbl}
\tcbuselibrary{skins}

\newcolumntype{Y}{>{\raggedleft\arraybackslash}X}

\tcbset{tab1/.style={fonttitle=\bfseries\large,fontupper=\normalsize\sffamily,
colback=yellow!10!white,colframe=red!75!black,colbacktitle=Salmon!40!white,
coltitle=black,center title,freelance,frame code={
\foreach \n in {north east,north west,south east,south west}
{\path [fill=red!75!black] (interior.\n) circle (3mm); };},}}

\tcbset{tab2/.style={enhanced,fonttitle=\bfseries,fontupper=\normalsize\sffamily,
colback=yellow!10!white,colframe=red!50!black,colbacktitle=Salmon!40!white,
coltitle=black,center title}}

\begin{document}
\title{Self-Tuning Spectral Clustering for Adaptive Tracking Areas Design in 5G Ultra-Dense Networks}
\author{Brahim Aamer$^{(1)}$, Hatim Chergui$^{(1,2)}$, Nouamane Chergui$^{(3)}$, Kamel Tourki$^{(4)}$, \\Mustapha Benjillali$^{(2)}$, Christos Verikoukis$^{(5)}$ and M\'erouane Debbah$^{(4)}$\\
{\normalsize{}$^{(1)}$ INWI, Casablanca, Morocco}\\
{\normalsize{}$^{(2)}$ Communication Systems Department, INPT, Rabat, Morocco}\\
{\normalsize{}$^{(3)}$ Huawei Technologies, Casablanca, Morocco}\\
{\normalsize{}$^{(4)}$ Huawei Tehnologies, Paris Research Center, France}\\
{\normalsize{}$^{(5)}$ CTTC, Barcelona, Spain}\\
{\normalsize{}Contact Email: \texttt{brahim.aamer@inwi.ma}}}
\maketitle
\thispagestyle{empty}
\begin{abstract}
In this paper, we address the issue of automatic tracking areas (TAs) planning in fifth generation (5G) ultra-dense networks (UDNs). By invoking handover (HO) attempts and measurement reports (MRs) statistics of a 4G live network, we first introduce a new kernel function mapping HO attempts, MRs and inter-site distances (ISDs) into the so-called similarity weight. The corresponding matrix is then fed to a self-tuning spectral clustering (STSC) algorithm to automatically define the TAs number and borders. After evaluating its performance in terms of the $Q$-metric as well as the silhouette score for various kernel parameters, we show that the clustering scheme yields a significant reduction of tracking area updates and average paging requests per TA; optimizing thereby network resources.
\end{abstract}


\section{Introduction}
\IEEEPARstart{A}{key} component in wireless networks is user location management. Such a function is achieved using the concept of tracking area (TA); similarly to location area (LA) in GSM and routing area (RA) in GPRS. To track users, a mobility management entity (MME) records the TA in which each user is registered. We consider TA design of cells managed by a single MME. When a user moves into a new TA, an update message is sent to
the MME. This causes a signaling overhead, referred to as
the update overhead. A second type of signaling overhead
exists in the reverse direction. In order to place a call to a
user, MME broadcasts a paging message in all cells of the
TA in which the user is currently registered. Having TAs
of very small size virtually eliminates paging, but leads to
excessive update, whereas very large TAs give the opposite
effect \cite{MME}.
\begin{figure}
\centering
\includegraphics[scale=0.65]{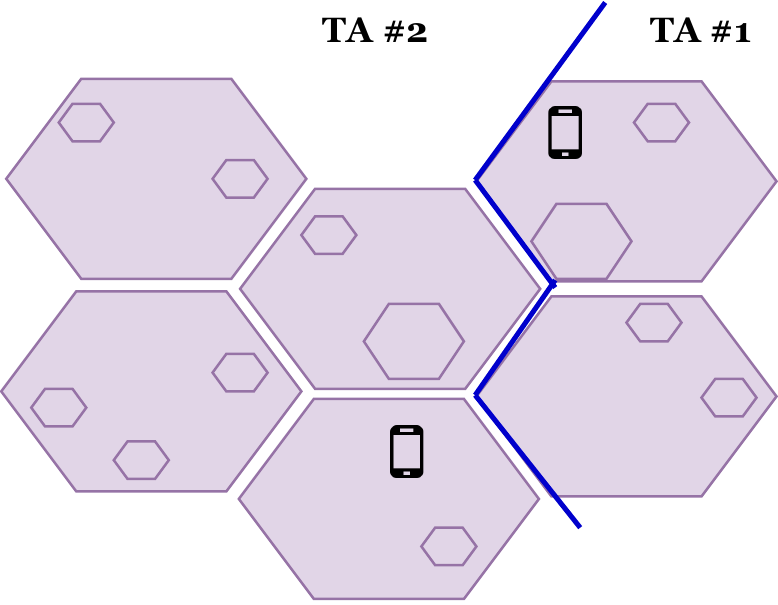}
\caption{Tracking areas are adopted to manage user location.}
\vspace{-0.5cm}
\end{figure}
\begin{figure*}[!ht] 
\normalsize 
\setcounter{MYtempeqncnt}{\value{equation}} 
\setcounter{equation}{0} 
\vskip -0.4cm
\begin{equation} 
s_{i,j} = \exp\left[-\left(\alpha\frac{d_{ij}}{\max\left(d_{ij}\right)}+\left(1-\alpha\right)\left(1-\beta\frac{a_{ij}}{\max\left(a_{ij}\right)}-\left(1-\beta\right)\frac{m_{ij}}{\max\left(m_{ij}\right)}\right)\right)\right]\label{similarity_matrix}  
\end{equation} 
\setcounter{equation}{\value{MYtempeqncnt}} 
\hrulefill 
\vspace{-5mm}
\end{figure*}
\begin{table*}[!htb]
\label{Table1}
\centering	
\newcolumntype{M}[1]{>{\centering\arraybackslash}m{#1}}

\caption{Samples from the Live Network Dataset}
\begin{tabular}{|M{2cm}|M{2cm}|M{2cm}|M{2cm}|M{2cm}|M{2cm}|}
\hline 
\multicolumn{1}{|>{\centering\arraybackslash}M{2cm}|}{\cellcolor{black!20} Source Site ID} & \multicolumn{1}{>{\centering\arraybackslash}M{2cm}|}{\cellcolor{black!20} Target Site ID} &\multicolumn{1}{>{\centering\arraybackslash}M{2cm}|}{\cellcolor{black!20} ISD (Km)}&\multicolumn{1}{>{\centering\arraybackslash}M{2cm}|}{\cellcolor{black!20} HO Attempts (Two Directions)}&\multicolumn{1}{>{\centering\arraybackslash}M{2cm}|}{\cellcolor{black!20} MRs Count}&\multicolumn{1}{>{\centering\arraybackslash}M{2cm}|}{\cellcolor{black!20} Paging Requests for Source Site}\\
\hline 
$45$& $45$ & $0$ & $30\,520$ & $161\,194$ & $79\,274$\\
$45$ & $294$ & $0.428617$ & $7\,275$ &  $108\,443$& $79\,274$\\
$56$ & $298$ & $0.66769$ & $1\,968$ &  $113\,536$ & $97\,725$\\
\hline
\hline
\end{tabular}
\vspace{-0.5cm}
\end{table*}
Intuitively, an optimal TA design tends to group new radio node Bs (gNBs) having
large numbers of users roaming between them, e.g., gNBs
along a road with much traffic, into the same TA. Nonetheless, the dynamic user behavior and traffic patterns in urban environments make that TAs, initially optimized for certain user statistics (or forecasts), become inaccurate and therefore urge to implement machine learning-driven adaptive TA design algorithms as part of the so-called self-organizing networks (SON) framework \cite{SON}.

\subsection{Related Work}
In \cite{Razavi1}, the authors presented a re-optimization approach
for revising a given TA design, which translates into a $NP$-hard optimization problem solved via repeated local search. Also, they presented in \cite{Razavi2} a ``rule of thumb'' method to allocate and assign tracking areas lists (TALs) for a network and compare the performance of an optimum conventional TA design with the suggested TAL design for a large scale network in Lisbon, Portugal. The results clearly showed the ability of dynamic TAL in reducing the signaling overhead and maintaining a good performance due to
reconfiguration compared to the conventional TA design. On the other hand, the performance of 4G TAL-based location management has been analyzed in \cite{Deng} using a Markov chain approach. The provided closed-form formulas highlight the effect of different network parameters on the signaling cost. In particular, the total signaling cost has been shown to be a downward convex function of the radius of a TA in terms of cells. In \cite{Toril}, the authors resorted to $K$-means clustering algorithm to perform automatic TA planning. This approach ensures the adaptation of the network to the changing user trends. Once a TA re-plan has been triggered, a graph partitioning algorithm is used to build the new TA plan.

\subsection{Contributions}

In this paper we investigate the following aspects:
\begin{itemize}
\item First, we construct a dataset featuring source/target sites relations, inter-site distances (ISDs), bidirectional HO attempts, events A3 MRs count and paging requests per source site. These features stem from automatic neighbor relation (ANR) statistics retrieved from the SON platform of a large 4G live network.

\item We introduce a new Gaussian kernel function that involves the three aforementioned features, and we use it to define an inter-site similarity matrix that is fed to the self-tuning spectral clustering (STSC) algorithm.

\item We show that the presented algorithm leads to the reduction of tracking area updates and average paging requests per TA, which optimizes radio network resources while automating TA design.

\end{itemize}

\vspace{-0.5cm}
\section{Dynamic TA Design Algorithm}
\subsection{Live Network Dataset}
The dataset is retrieved from the performance monitoring platform of a commercial 4G network. It specifies---for each couple of source-target sites---the daily bidirectional HO attempts as well as A3 event measurement reports (MRs). It also includes the paging requests per source site. The sites global positioning system (GPS) coordinates are used to calculate inter-site distances. The total number of measured neighboring relations is $N=149769$ corresponding to $M=387$ sites. Note that the adopted quantities, i.e., measurement reports and HO attempts are still viable in 5G.
\vspace{-0.3cm}
\subsection{Similarity Matrix}
A clustering algorithm operates on the so-called \textit{similarity matrix} whose entries measure the logical correlations between each couple of the dataset samples. In this regard, we adopt a radial basis function (RBF)-based precomputed matrix $\mathbf{S}$ that involves the three features, namely, the ISD, HO attempts and MRs count. The kernel parameter \texttt{gamma} is set to $1$. As such, the $(i,j)$th matrix element is given by Eq. (1) on top of this page, where $d_{ij}$, $a_{ij}$ and $m_{ij}$ stand for the pairwise distance, handover attempts and MRs count from site $i$ to site $j$. The parameters $\alpha,\,\beta\in\left[0,1\right]$ are controlling the dependency to each feature. The features are normalized with respect to their maximum. Hence, sites with low ISD or high attempts/MRs present a similarity weight near to $1$.
\setcounter{equation}{1}

\subsection{Self-Tuning Spectral Clustering}
As new cells are added and removed every day, we target a fully adaptive TA design algorithm that can process a high number of observations as well. In this respect, since basic clustering algorithms generally require that the number of clsuters be specified in the input, we resort to the well-established self-tuning spectral clustering (STSC) scheme. It was initially introduced in \cite{Self_Tuning}, where the authors studied a number of then open issues in spectral clustering:
\begin{itemize}
\item Selecting the appropriate scale of analysis, i.e., the parameter \texttt{gamma} of the RBF kernel,
\item Handling multi-scale data,
\item Clustering with irregular background clutter,
\item Finding automatically the number of clusters.
\end{itemize}
\begin{figure*}[t]
\centering
\subfloat[Live network TA design.]{\includegraphics[scale=0.45,trim={1cm 9cm 0cm 9cm},clip]{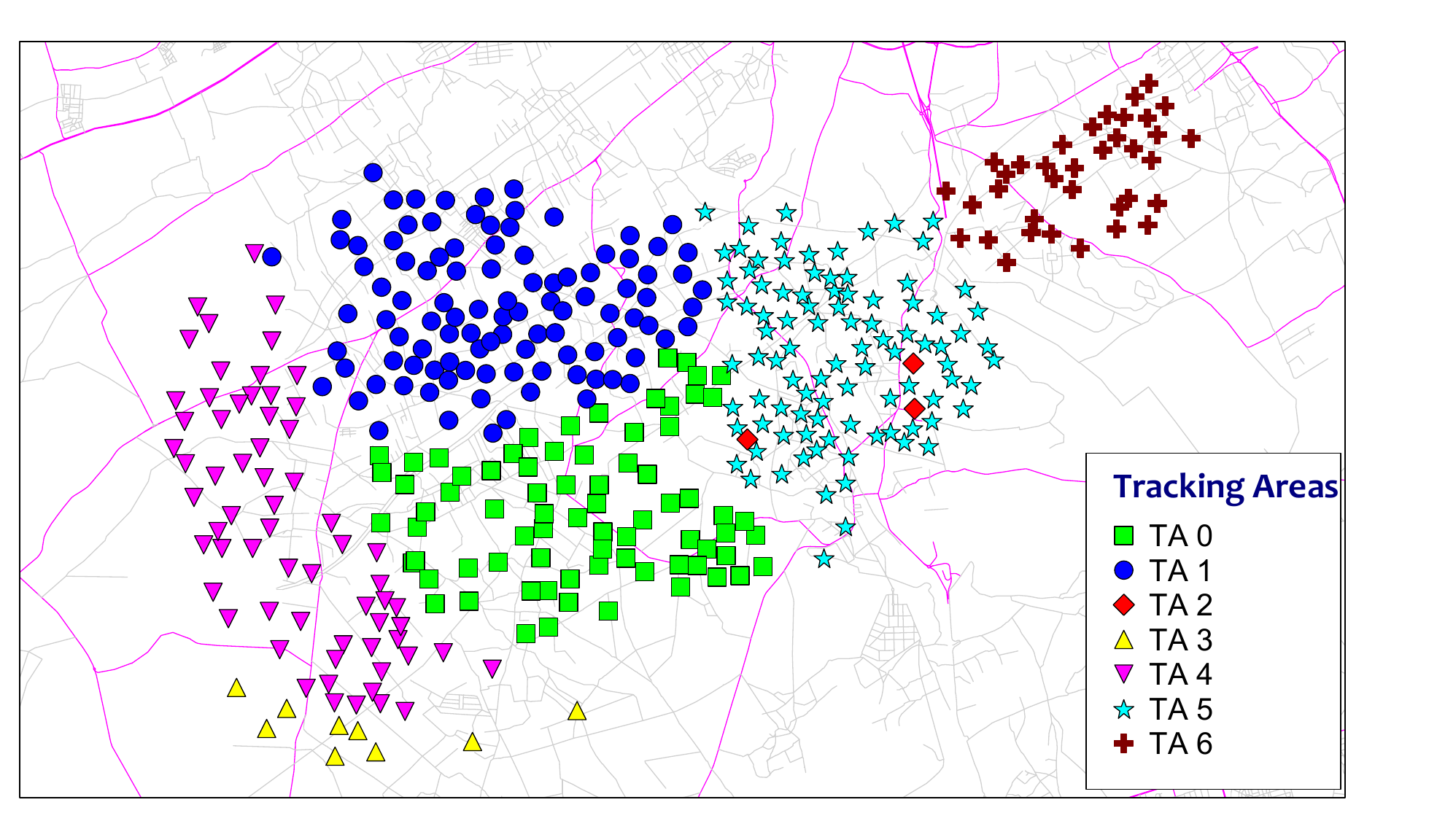}}
~~~~~~~~~~~\subfloat[$\alpha=0.4$ and $\beta=0.8$, STSC-based TA design.]{\includegraphics[scale=0.45,trim={1cm 9cm 0cm 9cm},clip]{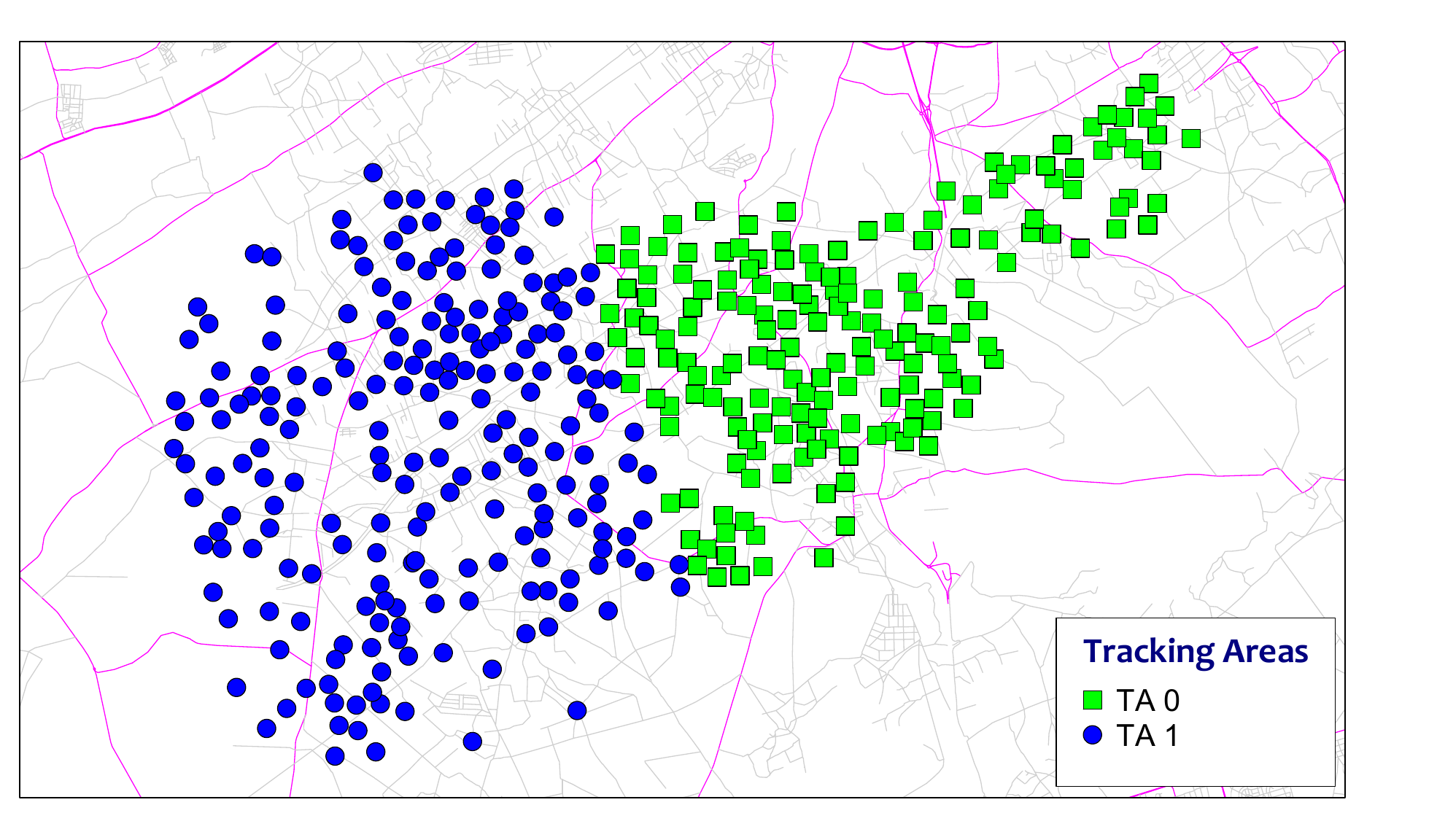}}\\
\subfloat[$\alpha=0.4$ and $\beta=0.5$, STSC-based TA design.]{\includegraphics[scale=0.45,trim={1cm 9cm 0cm 9cm},clip]{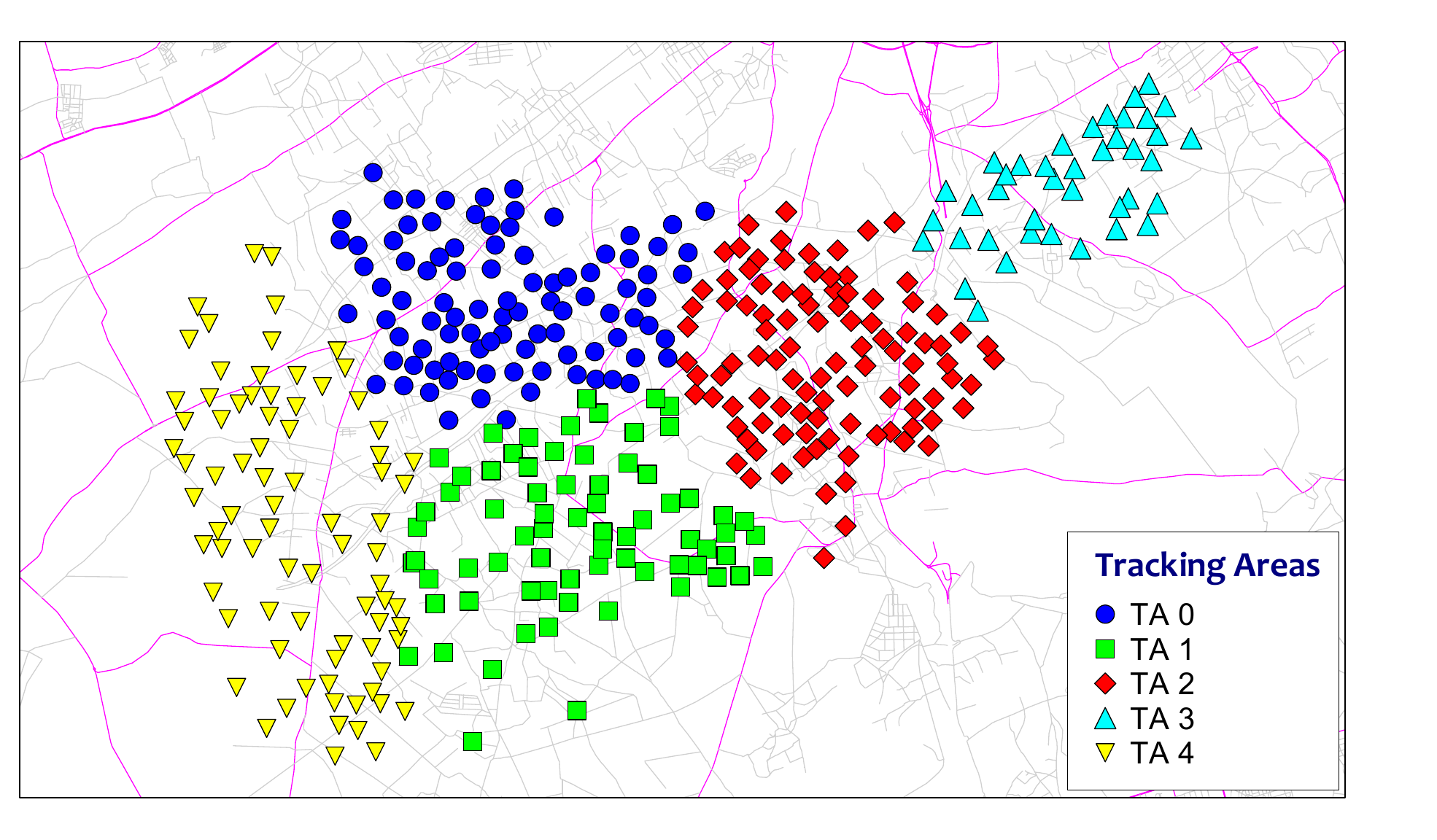}}
~~~~~~~~~~~\subfloat[$\alpha=0.5$ and $\beta=0.5$, STSC-based TA design.]{\includegraphics[scale=0.45,trim={1cm 9cm 0cm 9cm},clip]{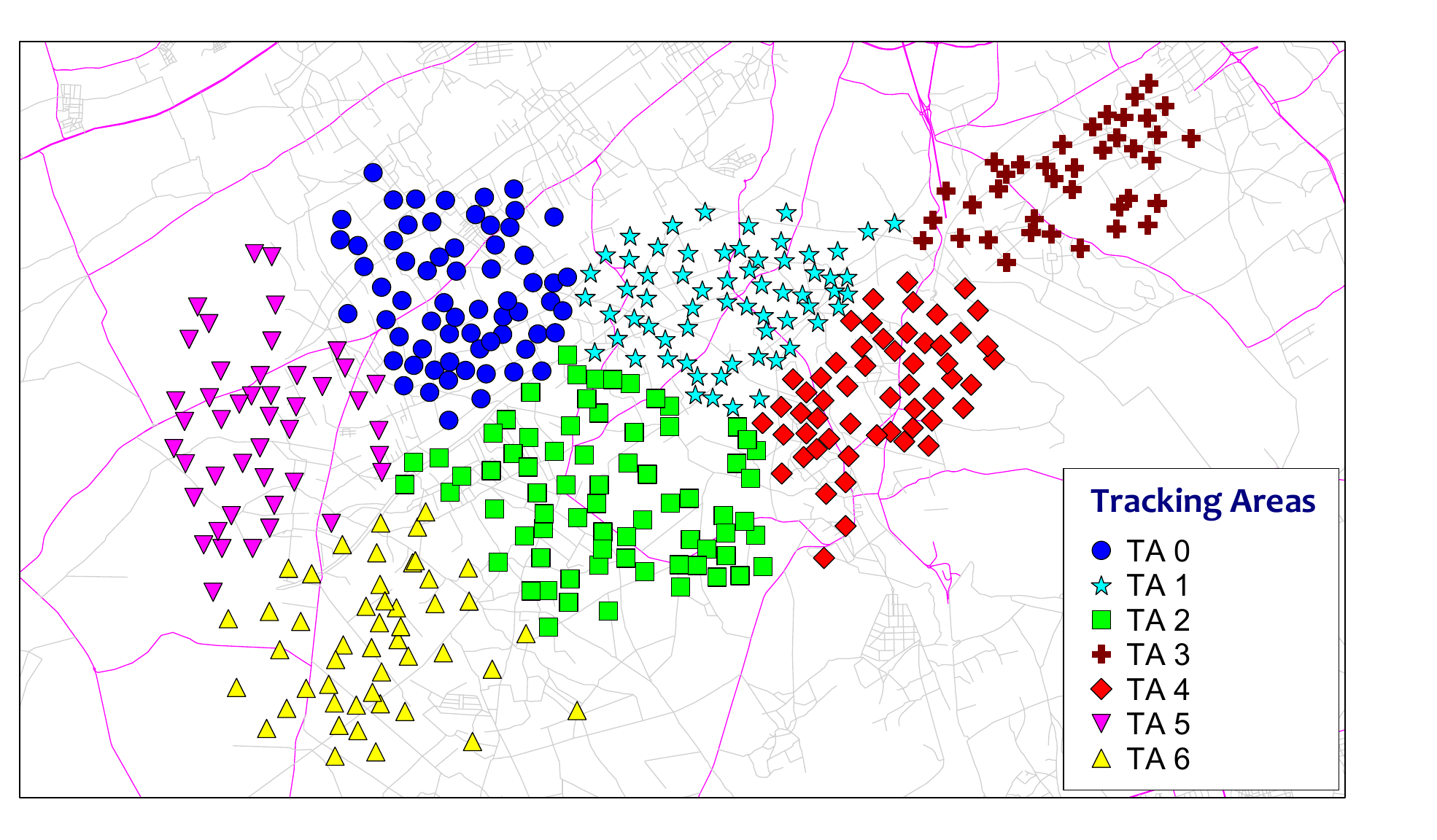}}
\caption{Adaptive TA design using self-tuning spectral clustering (STSC). For confidentiality considerations, the sites are displayed according to their shifted GPS coordinates.}
\end{figure*}
Spectral clustering algorithms \cite{Spectral_Clustering}, by definition, use the eigenvalues and eigenvectors (i.e., the spectrum) of the similarity matrix to cluster a dataset. After normalizing matrix $\mathbf{S}$, STSC uses the corresponding $C$ largest eigenvectors that are then stored in a matrix $\mathbf{X}$, where the first column of $\mathbf{X}$ is the biggest eigenvector.
One of the tasks of STSC is to get a cost for each possible number of clusters in the dataset to find the most probable
number of groups. This computation is done incrementally.
One starts with the minimum number of possible clusters $c_{\min}$ by taking the first $c_{\min}$ columns
of $\mathbf{X}$, we rotate them by applying a Givens rotation in an stochastic gradient descent scheme \cite{Self_Tuning} to find the optimal rotation matrix $\mathbf{R}$. The point of comparison when doing the rotation is the minimization of the cost function $J$ given by,
\begin{equation}
J = \sum_{i=1}^{M}\sum_{j=1}^{C}\frac{z_{i,j}^{2}}{\mu_{i}^{2}},
\end{equation}
with $\mathbf{Z}=\mathbf{X}\mathbf{R}$ and $\mu_{i}=\max_{j}z_{i,j}$. The parameters are updated in the opposite direction of the gradient $\nabla J$ following a learning rate. To that end, the STSC implementation \cite{Code} adopts automatic differentiation packages \texttt{autograd} \cite{Autograd} and \texttt{pymanopt} \cite{Pymanopt} to implement gradients.

The adopted STSC algorithm is detailed in Algorithm 1, and the selected parameters are listed in Table I.
\begin{table}[!hb]
\vspace{-4mm}
\label{Table2}
\centering	
\newcolumntype{M}[1]{>{\centering\arraybackslash}m{#1}}

\caption{Spectral Clustering Parameters}
\begin{tabular}{m{5cm}M{2.5cm}}
\hline 
\multicolumn{1}{>{\centering\arraybackslash}M{5cm}}{\cellcolor{black!20} Parameter} & \multicolumn{1}{>{\centering\arraybackslash}M{2.5cm}}{\cellcolor{black!20} Value}\\
\hline 
\texttt{kernel} & \texttt{rbf}\\
\texttt{gamma} & \texttt{1}\\
\texttt{Automatic differentiation} & \texttt{autograd}\\

\hline
\hline
\end{tabular}
\end{table}
\begin{algorithm}
  \caption{Adaptive Tracking Areas Design}
  \textbf{Inputs:} A dataset of sites $P=\left\{p_1,\ldots,p_M\right\}$ to be clustered. \\
  \textbf{Outputs:} Tracking area labels for the sites.
 
  \begin{algorithmic}[1]
   \STATE Construct a RBF-based similarity matrix $\mathbf{S}$ \\according to (\ref{similarity_matrix}).
   \STATE Set the optimization method to \texttt{autograd}.
   \STATE Define $\mathbf{D}$ to be the diagonal matrix with elements $\delta_{i,i}=\sum_{j=1}^{M}s_{i,j}$ and normalize matrix $\mathbf{S}$ as $\mathbf{L}=\mathbf{S}\mathbf{D}^{-1}$. 
   \STATE Initialize STSC \cite{Code} with the normalized similarity \\matrix $\mathbf{L}$.
   \STATE STSC yields an array of vectors corresponding to \\the clusters.
   \STATE Assign a TA label to the elements of each output \\vector. 
  \end{algorithmic}
\end{algorithm}
\begin{table*}[!t]
\vspace{-4mm}
\label{Table1}
\centering	
\newcolumntype{M}[1]{>{\centering\arraybackslash}m{#1}}

\caption{TAU and Paging Performance Gains}
\begin{tabular}{|M{1cm}|M{1cm}|M{2cm}|M{2cm}|M{2cm}|}
\hline 
\multicolumn{1}{|>{\centering\arraybackslash}M{1cm}|}{\cellcolor{black!20} $\alpha$} & \multicolumn{1}{>{\centering\arraybackslash}M{1cm}|}{\cellcolor{black!20} $\beta$} &\multicolumn{1}{>{\centering\arraybackslash}M{2cm}|}{\cellcolor{black!20} Number of TAs}&\multicolumn{1}{>{\centering\arraybackslash}M{2cm}|}{\cellcolor{black!20} TAUs}&\multicolumn{1}{>{\centering\arraybackslash}M{2cm}|}{\cellcolor{black!20} Paging Requests}\\
\hline 
\multicolumn{2}{|c|}{Live Network}  & $7$ & $1\,413\,898$ & $684\,569\,043$\\
{\cellcolor{Lavender}$0.5$} & {\cellcolor{Lavender}$0.5$} & {\cellcolor{Lavender}$7$} & {\cellcolor{Lavender}$1\,302\,330$} &  {\cellcolor{Lavender}$477\,110\,605$} \\
$0.3$ & $0.7$ & $6$ & $1\,701\,626$ &  $654\,860\,531$ \\
$0.4$ & $0.5$ & $5$ & $1\,072\,844$ &  $971\,173\,567$ \\
$0.3$ & $0.5$ & $4$ & $593\,046$ &   $1\,432\,542\,174$ \\
$0.7$ & $0.3$ & $3$ & $726\,496$ &   $2\,544\,265\,641$\\
$0.4$ & $0.8$ & $2$ & $177\,554$ &   $5\,702\,369\,104$\\ 

\hline
\hline
\end{tabular}
\vskip -0.3cm
\end{table*}

\section{Performance Assessment}
In this section, we compare our STSC-based TA planning with the live network manual TA design depicted in Fig.~2-(a). The latter consists of 7 tracking areas with non-uniform sites distribution. Moreover, it includes an inaccurately planned TA $2$. Before delving into the qualification of our clustering algorithm, let us study its convergence across different settings of parameters $\alpha$ and $\beta$ and assess the potential gains in terms of tracking area updates (TAUs) and paging requests.

\begin{figure}[h!]
\vspace{-0.5cm}
\centering
\includegraphics[scale=0.55]{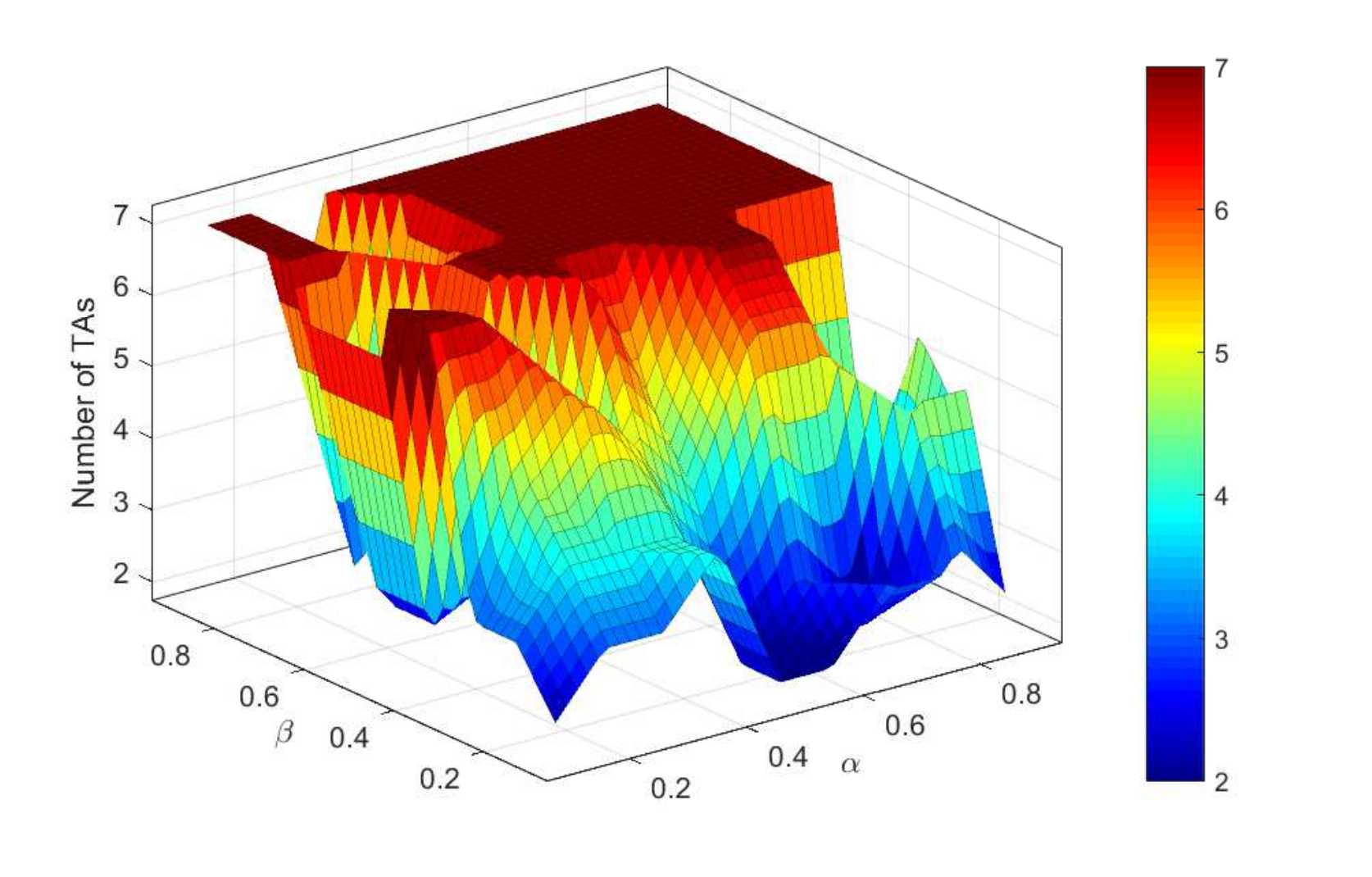}
\caption{Number of TAs vs. parameters $\alpha$ and $\beta$.}
\vspace{-0.5cm}
\end{figure}
\subsection{Number of Clusters}
A key remark, revealed by Fig.~2 and confirmed by Fig.~3, is that STSC converges to $7$ clusters when $\alpha$ and $\beta$ are greater or equal to $0.5$, i.e., when ISD---on one hand---and the mixture of HO attempts and MRs count---on the other hand---are fairly taken into account in the construction of the similarity matrix. This is also a reasonable criteria to avoid clustering distant overshooting sites just because their MR count is high, or e.g., grouping close sites with low HO events. Under this setting, the STSC algorithm finds a trade-off between minimizing average paging requests per cluster and reducing the inter-cluster HO attempts and thereby tracking area updates. In contrast, relying on the ISD only does not consider the clutter effect that is more pronounced in the HO attempts and MRs count.  

In practice, we note that for $\alpha \geq 0.5$, the number of clusters increases linearly with $\beta$. In this regime, radio network planning engineers may fix $\alpha=0.5$ and control the number of output TAs by fine-tuning $\beta$.

\subsection{TAU and Paging Performance Gains}
Table III depicts the TA updates and average paging requests per TA for various combinations of $\alpha$ and $\beta$. In this regard, we remark that---as expected---increasing the number of clusters leads to the augmentation of TAUs and decrease of paging requests per TA. With $7$ clusters, our STSC-based scheme achieves a better performance compared with the live network, with a reduction of $8\%$ in TAUs and $30\%$ in paging requests. In practice, radio network planning engineers may fine-tune $\alpha$ and $\beta$ to control the load of either the tracking area updates or paging in such a way to optimize radio resources (e.g., downlink (DL) physical resource blocks (PRBs) consumed by the S1 paging).
\begin{figure}[h!]
\centering
\includegraphics[scale=0.55]{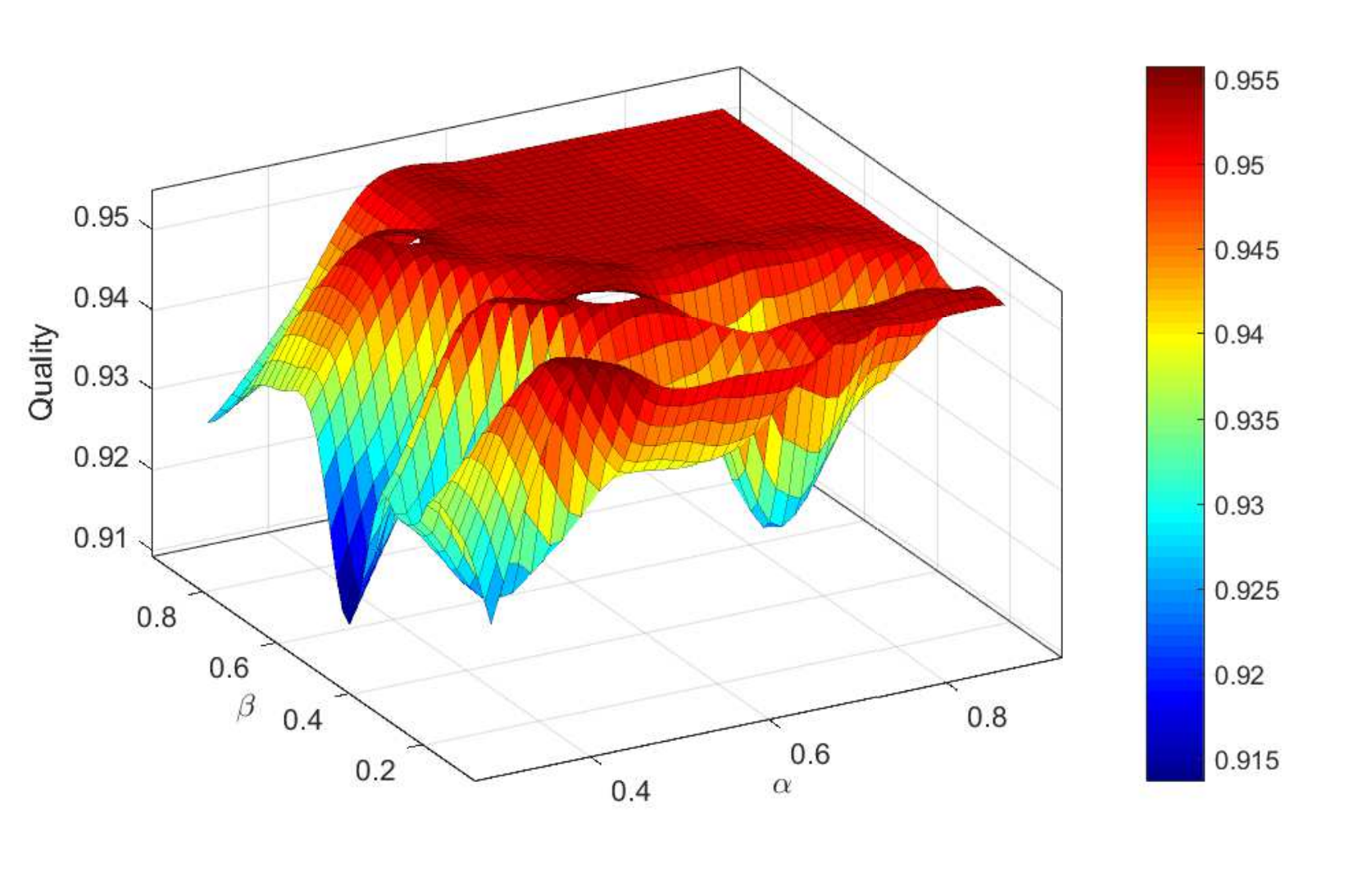}
\caption{STSC clustering quality vs. parameters $\alpha$ and $\beta$.}
\end{figure}
\subsection{STSC Quality}
Upon the convergence of STSC, the minimum cost $J_{\min}$ is used to define the quality $Q$ of the clustering with values ranging between $0$ and $1$ \cite{Self_Tuning}, where
\begin{equation}
Q = 1 - \frac{(J_{\min}/M)-1}{C}.
\end{equation}
A high quality clustering means that almost every site is assigned to the closest cluster, minimizing thereby the cost function $J$ and approaching to $Q=1$. In contrast, rural sites---that usually present a sparse distribution---increase the cost function and reduce the global clustering quality. In this respect, STSC generally groups suburban sites in the same TA. This is the case of TA $3$ in Fig.~2-(c).

In Fig.~4, we plot the $Q$-metric versus $\alpha$ and $\beta$, and we notice that the best quality is obtained when $\alpha,\,\beta\in\left[0.5,1\right]$, i.e., when the number of clusters is $7$.

\subsection{Silhouette Score}
The silhouette score $\sigma$ displays a measure of how close each point in one cluster is to the points in the neighboring clusters, and thus provides a way to assess parameters like number of clusters visually. This measure has a range of $\left[-1, 1\right]$ and is readily available in \texttt{scikit-learn} package \cite{Scikit}. The corresponding expression for site $i$ reads,
\begin{equation}
\sigma_i = \frac{\bar{d}_i-d_i}{\max\left(\bar{d}_i,d_i\right)},
\end{equation}
where $d_i$ and $\bar{d}_i$ stand for the average distance of site $i$ to the sites within the same cluster and the smallest average distance of $i$ to all sites in any other cluster, respectively.
The obtained positive silhouette score in Fig.~5 means that---in average---the dataset sites are assigned to a close cluster (in terms of the similarity weight defined in (\ref{similarity_matrix})). Note that sparsely distributed suburban sites such as TA $3$ in Fig.~2-(c) might present generally a negative silhouette score that degrades the overall performance.

\begin{figure}[t]
\centering
\includegraphics[scale=0.55]{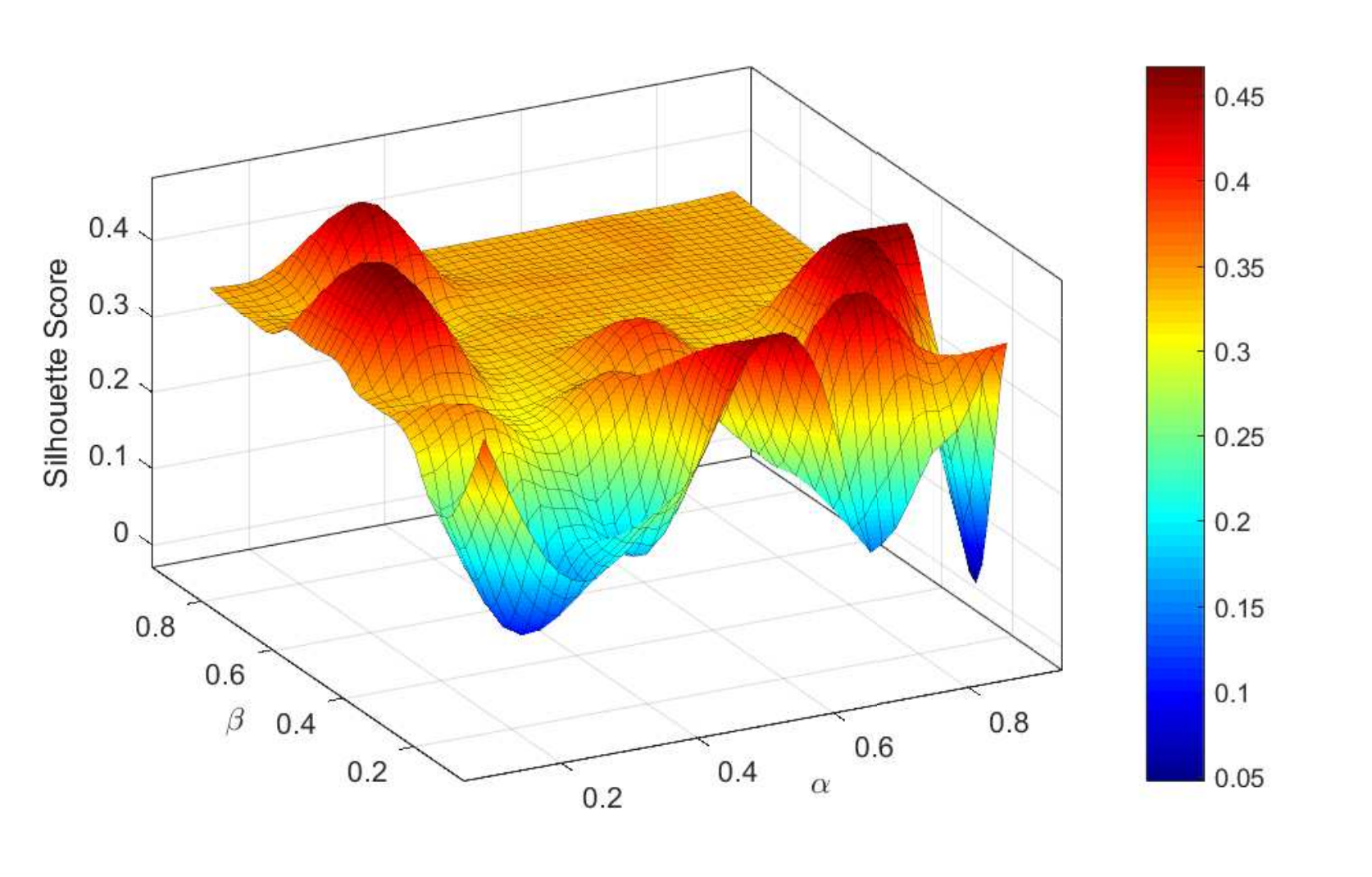}
\caption{Silhouette score vs. parameters $\alpha$ and $\beta$.}
\end{figure}

\vspace{1cm}
\section{Conclusion}
In this paper, we have presented a machine learning approach to automate tracking areas design in future 5G networks. It relies on a self-tuning spectral clustering algorithm capable of grouping gNBs without requiring the number of clusters as input. Alternatively, we feed STSC with a new kernel similarity matrix; taking into account inter-site distance, handover attempts and A3 events measurement reports count. The presented approach yields a significant reduction of tracking area updates and average paging requests per TA, and might be adopted by radio network planning engineers to periodically update TA design according to network evolution.

\section*{Acknowledgement}
This work is supported by the telecom operator INWI, Casablanca, Morocco.

\balance

\end{document}